\documentclass[a4paper, 12pt]{article}
\usepackage{a4}
\usepackage{amsfonts}
\usepackage{amssymb}
\usepackage{epsfig}
\usepackage{psfig}

\newcommand{\be}{\begin{equation}}
\newcommand{\ee}{\end{equation}}
\newcommand{\bea}{\begin{eqnarray}}
\newcommand{\eea}{\end{eqnarray}}
\newcommand{\mbb}{\mathbb}

\begin{document}
\linespread{1.2}

\title{
\begin{flushright} \vspace{-2cm} 
{\small DAMTP-2004-85 \\ \vspace{-0.35cm} 
hep-th/0409215} \end{flushright}
\vspace{3cm}
{\bf On the Explicit Construction   and  Statistics of Calabi-Yau Flux Vacua}
}
\author{} 
\date{}

\maketitle

\begin{center}
Joseph P. Conlon \footnote{e-mail: J.P.Conlon@damtp.cam.ac.uk}
and
Fernando Quevedo \footnote{e-mail: F.Quevedo@damtp.cam.ac.uk} \\
\vspace{0.5cm}
\emph{DAMTP, Centre for Mathematical Sciences,} \\
\emph{Wilberforce Road, Cambridge, CB3 0WA, UK} \\
\end{center}

\begin{abstract}
\noindent
We explicitly construct and study the  statistics of flux vacua 
for type IIB string theory 
on an orientifold of the Calabi-Yau hypersurface
$\mbb{P}^4_{[1,1,2,2,6]}$,  parametrised by two relevant
complex structure moduli. We solve for these moduli and the
dilaton field in terms of the set of integers defining the $3$-form
fluxes
and examine the distribution of vacua.
 We  compare our numerical results 
with the predictions of the Ashok-Douglas
density $\det (-\mathcal{R} - \omega)$, finding good overall agreement
 in different regions of moduli space. 
The number of vacua are found to scale with the distance
in flux space.
Vacua cluster in the
 region close to
the conifold singularity. Large supersymmetry breaking is more generic
but supersymmetric and hierarchical supersymmetry breaking vacua can
also be obtained. In particular, the small superpotentials and large
dilaton VEVs
needed to obtain de Sitter space in  a controllable approximation are possible but
not generic.  We argue that
in a general flux compactification, the rank of the gauge group coming
from D3 branes could be statistically preferred to be very small.
\end{abstract}

\thispagestyle{empty}
\clearpage

\tableofcontents

\section{Introduction}
\linespread{1.2}

It has been  claimed that the current status of string theory resembles that of
particle physics in the early 1960's\cite{joep} in the sense that there are many
experimental results but no organising principle.
 For string theory the place of the experiments
is taken by the many explicit string vacua that have been constructed
over the years. One desirable avenue towards
understanding string theory involves the development and questioning 
of its basic principles. Another less ambitious but equally important
program is to explore the large variety of string vacua with the
hope of learning more about the structure of the theory and of finding some
of its phenomenological and/or cosmological implications. 

In the phenomenological direction there has been substantial progress
through the construction of string vacua with properties very similar to
the standard model\cite{revphen}. In cosmological applications,
 concrete ways have been found to realise inflation within string
theory \cite{revinflation}. Furthermore the construction and study of
string vacua and their properties
has played an important role over the years in the understanding of
the theory itself, as exemplified by the discovery of $T$-duality \cite{tduality},
mirror symmetry \cite{mirror} and $S$-duality \cite{sduality}.
 Mirror symmetry provides probably the most
direct example in the sense that even though there were several ideas
about its existence, the actual realisation came only after a plot of
many of the vacua was found to be essentially symmetric under the exchange of
complex and K\"ahler structure moduli, thus providing `experimental'
evidence for this symmetry.

Whether we like it or not, the lack of a
close experimental test of the theory and the large abundance of vacua
suggest that a reasonable direction to pursue is to create a large
database of string vacua as a temporary substitute for experimental
results. This database may eventually be used to extract
model-independent features of string vacua that may be testable. It may also
give us some further hints into the structure of the theory itself.

These issues have been recently clarified through the study of antisymmetric
tensor flux
compactifications \cite{fluxes, gkp}.
In the absence of these fluxes there were myriads of vacua, each
with a large continuous degeneracy parametrised by the moduli of
the compactification space. Turning on the fluxes substantially reduces
this degeneracy by making it discrete, owing to the Dirac quantisation
condition for the fluxes. The fluxes can fix the values of
the moduli, which then belong to a discrete set
determined by the allowed fluxes. This huge discrete degeneracy 
makes a statistical analysis suitable for the study of vacua.
The stabilisation of the dilaton and complex structure moduli is
relatively straightforward, but
flux compactifications may also enable the construction of 
metastable vacua with positive
cosmological constant with the K\"ahler structure moduli also
stabilised as in the KKLT scenario \cite{kklt, desitterev}.

Even though the number of flux vacua appears to be enormous for a 
general Calabi-Yau manifold, there are a number of requirements that must
be satisfied for these vacua to be acceptable. First, as the entire
framework is founded on a weak coupling approximation, 
only solutions leading to values of $\rm{Im } \, \tau \gg 1$, where 
$\tau = C_0 + i e^{-\phi}$ is the dilaton-axion field that measures the string coupling.
The self-duality of type IIB under the $SL(2,\mbb{Z})$
$S$-duality symmetry implies that any solution can be
mapped to the fundamental domain.
Having mapped solutions to this region we can then only trust values of the dilaton
substantially greater than one. As we will discuss in section \ref{fpm}, this is expected to be naturally
suppressed statistically, with the proportion of vacua with $\rm{Im }
\, \tau > \tau_0$ falling as $\frac{1}{\rm{Im } \, \tau_0}$.
Secondly, the fluxes may or may not break supersymmetry by
themselves and it is of interest to know the value of the supersymmetry
breaking scale for each model. This is more urgent if the models are
embedded in the full KKLT scenario in which non-perturbative
superpotentials are also included to fix the K\"ahler
moduli\footnote{In the KKLT scenario, the combination of fluxes that themselves break
  supersymmetry with a non-perturbative superpotential results in a
  supersymmetric minimum. In this case
  supersymmetry is broken only after the introduction of anti
  D3-branes \cite{kklt,egq}, magnetic fluxes on D7 branes \cite{bkq},
  by $\alpha'$ corrections \cite{hepth0408054},
or in other vacua of the complex structure moduli\cite{eva}.}. In order to
guarantee a volume sufficiently larger than the string scale and
believe the four-dimensional effective field theory analysis, the value of the
gravitino mass induced only by the fluxes has to be hierarchically
small, a condition that is not straightforward to satisfy.

Recently, Douglas and collaborators have developed statistical
techniques to study string vacua
e.g.\cite{hepth0303194, hepth0307049, hepth0404116}. In particular, in
\cite{hepth0307049}, a general formula was proposed that captures the
statistical nature of the vacua, 
in terms of a density given by $\det (-\mathcal{R} - \omega)$ where $\mathcal{R}$ is the
curvature
two form and $\omega$ the K\"ahler two-form on moduli space. In 
\cite{hepth0404243} the validity of this formula was
successfully tested in a one-modulus example.

In this note we extend this analysis to a simple two-modulus
compactification, corresponding to the three-fold realisable as a
hypersurface in the weighted projective space $\mbb{P}_{[1,1,2,2,6]}$.
We turn on 3-form fluxes and use the
techniques of Candelas {\it et al} \cite{quintic, general, twomod} to compute the periods. This allows
us to determine the superpotential for the dilaton and complex structure
moduli
fields. We use a Monte Carlo analysis to study the distribution of
vacua in the vicinities of the Landau-Ginzburg and conifold regions of
moduli space. 
We compute the Ashok-Douglas index and compare with
our results, finding good general agreement. We also investigate the
distribution of values of the dilaton field as well as the
supersymmetry breaking scale.

While in the last stages of this project we received an article
 \cite{mn} that discusses the Ashok-Douglas density
for the same Calabi-Yau without performing the numerical analysis.
 
\section{General Framework}
\label{Model}

We here review the construction of the particular Calabi-Yau we study, 
and the calculation of its periods\cite{general,twomod}. Some explicit flux
compactifications on this Calabi-Yau are also presented in \cite{hepth0312104}.

\subsection{The Model}

There exist many Calabi-Yau manifolds that can be realised as
hypersurfaces in weighted projective space. The weighted projective
space $\mbb{P}^4_{[k_0, k_1, k_2, k_3, k_4]}$ is defined by
$$
(w_0,w_1,w_2,w_3,w_4) \equiv (\lambda^{k_0} w_0, \lambda^{k_1} w_1, \lambda^{k_2} w_2,
\lambda^{k_3} w_3, \lambda^{k_4} w_4)
$$
and has four complex dimensions. To obtain a space of three complex
dimensions we restrict to the hypersurface $P(w_i) = 0$, where $P$ is
a polynomial in the $w_i$. The condition that such a hypersurface be
Calabi-Yau is that $\textrm{deg}(P) = \sum_{i=0}^4 k_i.$
In this paper we will make use of the Calabi-Yau hypersurface in
$\mbb{P}^4_{[1,1,2,2,6]}$, which arises from
\be
\label{polyeqn}
 w_0^{12} + w_1^{12} + w_2^{6} + w_3^{6} + w_4^{2} = 0.
\ee
We shall denote this manifold by $\mathcal{M}$.
$\mathcal{M}$ has $h^{1,1} = 2$ and $h^{2,1}
= 128$. The number of complex structure moduli is determined
by the number of monomial deformations of degree 12 that can be added
to (\ref{polyeqn}). There are two complex structure
moduli that are particularly important, which we shall denote by
$\psi$ and $\phi$. These perturb (\ref{polyeqn}) to
\be
\label{manifoldeq}
P(w_i) \ =\ w_0^{12} + w_1^{12} + w_2^{6} + w_3^{6} + w_4^{2}
-12\psi\,\left( w_0 w_1 w_2 w_3 w_4\right) - 2\phi\, \left( w_0^6 w_1^6\right) = 0.
\ee
Their significance is due to their survival in the mirror manifold $\mathcal{W}$,
obtained by identifying points in $\mathcal{M}$ related by the action
of $G$, the maximal group of scaling
symmetries of (\ref{polyeqn}). $G$ here is $\mbb{Z}_2 \times
\mbb{Z}_6^2$, and its action is represented by 
$$
\mbb{Z}_2:(w_1,w_4)  \to   (\alpha w_1, \alpha w_4)\textrm{ with } \alpha^2=1,
$$
$$\mbb{Z}_6:
(w_1,w_3)  \to  (\beta^5 w_1,  \beta w_3) \textrm{ with } \beta^6=1,$$
$$\mbb{Z}'_6:
(w_1,w_2)  \to  (\gamma^5 w_1, \gamma w_2) \textrm{ with } \gamma^6=1.$$
These manifestly leave equation (\ref{manifoldeq}) invariant. All
other degree 12 deformations of equation (\ref{manifoldeq}) 
are not well defined on the mirror. As $\mathcal{W}$ has $h^{1,1} = 128$ and $h^{2,1} = 2$,
it is an example of a Calabi-Yau with two complex structure
moduli. Both $\mathcal{M}$ and $\mathcal{W}$ develop a conifold
singularity when $864 \psi^6 + \phi = 1$. This condition follows from
requiring $P(w_i) = \underline{\nabla} P(w_i) = 0$ with
$\underline{\nabla} \, \underline{\nabla} \, P$ non-singular.

The Ashok-Douglas index density describes the statistics of
the class of models described in \cite{gkp}, namely flux
compactifications of type IIB orientifolds.
To fit the above model into this framework, we embed it into an $F$-theory model with an orientifold limit 
according to Sen's prescription\cite{hepth9702165}. The corresponding
elliptically fibered
four-fold is the hypersurface in weighted projective space 
$\mbb{P}^5_{[1,1,2,2,12,18]}$ with $\chi=19728$. The associated
  tadpole cancellation condition then reads\cite{hepth9906070}:
\be
N_{\rm D3}+N_{\rm flux} =\frac{\chi}{24}= 822\equiv L.
\ee
Here $N_{\rm D3}$ is the net number of D3 branes and 
$N_{\rm flux}$ the contribution of the fluxes to the D3 brane charge,
\be
N_{\rm flux}\ =\ \frac{1}{\left(2\pi\right)^4\alpha'^2}\, \int_{\cal
  M} H_3\wedge F_3.
\ee
Here $H_3 = dB_2$ and $F_3 = dC_2$ are the NS-NS and R-R 3-forms respectively.
To cancel tadpoles in a supersymmetric fashion we then require the fluxes to carry
at most 822 units of D3-brane charge (this number as usual measures the
  amount of negative D3 brane charge coming from D7 branes and
  orientifolds, in the orientifold realisation of the model).

In the low-energy approximation, the moduli of the
Calabi-Yau appear as scalar fields in a supergravity theory.
The effective four-dimensional ${\cal N}=1$ supersymmetric field theory is
determined  by the K\"ahler potential $K_T(\xi_p, \phi_i)$ and the
superpotential $W(\phi_i)$, where $\xi_p \left( p=1, \ldots, h^{1,1}
\right)$ represent the K\"ahler
moduli and $\phi_i \left( i =1, \ldots h^{2,1}+1 \right)$ the dilaton and complex structure moduli. The
tree-level K\"ahler potential takes the form
\be
K_T(\xi_p, \phi_i)= \hat K(\xi_p) \ + K(\phi_i),
\ee
where $\hat K$ is the K\"ahler potential for the K\"ahler moduli and
$K$ that for the dilaton and complex structure moduli. We will
write an explicit expression for $K$ in the next section. As for $\hat
K$, the only information we need to provide is that it is of the
no-scale form, in the sense that $\hat{K}{^{-1}}_p^{\phantom{1} q} \hat{K}^{p}\hat{K}_q=3$ 
where  $p$ and $q$ label the K\"ahler moduli.

The appropriate superpotential for non-vanishing fluxes was proposed
by
Gukov, Vafa and Witten \cite{hepth9906070} and takes the
form
\be
W\ =\ \int_{\cal M}\left (F_{3}- \tau H_{3}\right)\wedge \Omega.\ee
$\tau$ is the
complex dilaton-axion field and 
$\Omega$ the holomorphic $(3,0)$ form for the Calabi-Yau $\cal M$.

The combination of the fact that $W$ does not depend on the K\"ahler moduli and the
 no-scale structure of the K\"ahler potential
  implies that the effective
scalar potential is simply given by
\be V=e^{K_T}\left(D_i W  D_{\bar {j}}\bar{W} K^{-1}_{i\bar{j}}\right),\ee
where $D_iW=\partial_i W+(\partial_i K) W$ (in Planck mass units). 
Therefore, to stabilise these fields
at the minimum of the potential we need $D_i W=0$. The K\"ahler
moduli
only appear in the potential in the $e^{K_T}$ overall factor and are
not fixed. We will omit these fields in the rest of this paper except
for pointing out that the no-scale structure implies that the order
parameter for supersymmetry breaking is 
\be
D_p W\  =\ {(\partial_p \hat {K}) W}. \ee
Thus a non-vanishing value of the superpotential $W$ implies
supersymmetry breaking by the $F$-term of the K\"ahler structure
moduli. Since a K\"ahler transformation $K\to K+f+\bar f,\,  W\to e^{-f}
W$
leaves the action invariant, the value of $W$ can always be
rescaled and a more appropriate, 
 K\"ahler transformation invariant, measure of supersymmetry
breaking is the gravitino mass, given by
\be
m^2_{3/2}\ =\ e^K |W|^2. \ee

Notice that in the KKLT  construction, a non-perturbative
superpotential depending on these fields is added \cite{kklt,egq}, breaking the
no-scale structure and fixing the K\"ahler moduli. Anti D3-branes (or magnetic
fluxes on the D7-branes \cite{bkq}) are introduced in order to lift
the minimum of the potential to a positive  value. This can certainly be done in the
present model, but as we have nothing new to say in this respect 
we concentrate on the dependence of the theory on the
dilaton and complex structure moduli, to which we now turn.
 
\subsection{Fluxes, Periods and Moduli}
\label{fpm}

For any Calabi-Yau 3-fold, the middle homology and cohomology are naturally expressed in 
terms of a symplectic basis.
That is, there exists a basis of 3-cycles $A^a$ and $B_b$ and a
basis of 3-forms $\alpha_a$ and $\beta^b$ (where $a,b=1,2 \ldots (h^{2,1}+1)$), such that
in homology
\bea
A^a \cap B_b = -B_b \cap A^a & = & \delta^a_b, \nonumber \\
A^a \cap A^b = B_a \cap B_b & = & 0,
\eea
and
\bea
\int_{A^b} \alpha_a = - \int_{B_a} \beta^b & = & \delta^b_a ,\\
\int_{\mathcal{M}} \alpha_a \wedge \beta^b = -\int_{\mathcal{M}} \beta^b \wedge
\alpha_a & = & \delta^b_a .
\eea
Such a symplectic basis is only defined up to $Sp(2n,\mathbb{Z})$
transformations, as these preserve the symplectic intersection form.
The periods are defined as the integral of the holomorphic 3-form $\Omega$ over these cycles,
\be
\int_{A_a} \Omega = z^a,\qquad \qquad
 \int_{B^b} \Omega = \mathcal{G}_a.
\ee
The periods are encapsulated in the period vector, $\Pi =
\left( \mathcal{G}_1, \ldots ,\mathcal{G}_n,z_1,\ldots, z_n \right)$, 
where $n = h^{2,1} + 1$. This inherits the holomorphic freedom 
of $\Omega$ and is defined up to holomorphic rescalings $\Omega \to
f(\phi_i)\Omega$. Note that $\Pi = \Pi(\phi_i)$, 
where $\phi_i$ are the complex structure moduli of the Calabi-Yau.

Given the vector of periods $\Pi(\phi_i)$, the K\"ahler potential on complex
structure moduli space is given by \cite{px}
\bea
K(\tau, \phi_i) & = &  -\ln(-i(\tau - \bar{\tau})) - \ln \left( i \int
\Omega \wedge \bar{\Omega} \right) \nonumber \\
& = & -\ln(-i(\tau - \bar{\tau})) - \ln(-i\Pi^\dagger
\cdot \Sigma \cdot \Pi) \\
& \equiv & K_\tau + K_\phi. \nonumber
\eea
where
$$
\Sigma = \left( \begin{array}{cc} 0 & \mathbf{1}_n \\ -\mathbf{1}_n & 0 \end{array} \right)
$$
and $\tau$ is the dilaton-axion.
We can then compute the metric on moduli space,
\be
g_{\alpha \bar{\beta}} = \partial_\alpha \partial_{\bar{\beta}} K,
\ee
and the Riemann and Ricci curvatures
\bea
R^{\lambda}_{\mu \bar{\nu} \rho} & = & -\partial_{\bar{\nu}}
(g^{\lambda \bar{\alpha}} \partial_\mu g_{\rho \bar{\alpha}} ),
\nonumber \\
R_{\mu \bar{\nu}} = R^{\lambda}_{\mu \bar{\nu} \lambda} & = & - \partial_\mu \partial_{\bar{\nu}} \log (\det g).
\eea

In terms of the periods, the Gukov-Vafa-Witten superpotential is
\be
\label{gvwpot}
W = (2 \pi)^2 \alpha' (f - \tau h)\cdot \Pi(\phi_i),
\ee
where $f = (f_1, \ldots, f_6)$ and $h = (h_1, \ldots, h_6)$ are
integral vectors of fluxes along the cycles. The fluxes $F_3$ and
$H_3$ are elements of $H^3(\mathcal{M}, \mbb{Z})$ and satisfy the flux
quantisation conditions
\be
\frac{1}{(2 \pi)^2 \alpha'} \int_{\Sigma^3} F \in \mbb{Z}, \qquad 
\frac{1}{(2 \pi)^2 \alpha'} \int_{\Sigma^3} H \in \mbb{Z},
\ee
where $\Sigma^3$ is an arbitrary 3-cycle.
The amount
of $D3$-brane charge carried by the fluxes is then
\be
\label{Nflux}
N_{flux} = f^T \cdot \Sigma \cdot h.
\ee

The stabilisation of the complex structure moduli $\phi_i$ is governed
by the following equations
\bea
\label{fluxeqns}
D_{\phi_i} W = \partial_{\phi_i} W + (\partial_{\phi_i}K)W & = & 0, \nonumber\\
D_\tau W = \partial_{\tau} W + (\partial_{\tau}K)W & = & 0.
\eea
Distinct choices of fluxes stabilise the complex structure
moduli at discrete points in moduli space.
As the total number of flux vacua is very large, the distribution of
discrete flux vacua can be approximated by a continuous distribution.
Treating the fluxes as non-quantised, Ashok and Douglas\cite{hepth0307049} derived a formula
for an index that gives an estimate of the total number of vacua with
$N_{\mathrm{flux}} < L$ on a region $\mathcal{X}$ of moduli space
\be
\label{ashokdouglasdensity}
I_\mathrm{vac}(N_{\mathrm{flux}} \leq L) = \frac{(2 \pi L)^K (-1) ^{\frac{K}{2}}}
{\pi^{n+1} K!} \int_\mathcal{X} \det(-\mathcal{R} - \mathbf{1} \cdot \omega)
\ee
where $\omega$ is the K\"ahler form and $\mathcal{R}$ the curvature two-form on
moduli space.
In formula (\ref{ashokdouglasdensity}) $K$ is the number of cycles
along which flux is wrapped and $L$ the total available $D3$-brane charge.
In a comparison with numbers of explicit vacua, the constant prefactor is obviously not relevant. 

We will now restrict to cases where the Calabi-Yau $\mathcal{M}$ has
two complex structure moduli.
The region $\mathcal{X}$ includes the dilaton-axion moduli space. The 
weighted vacuum density over the Calabi-Yau moduli space is then evaluated
to be\cite{kgprivate}
\be
\label{indexDensity}
d \mu =  g_{\tau \bar{\tau}} d\tau \wedge d \bar{\tau} \wedge 
\left( 4 \pi^2 c_2 - \det(g_{a \bar{a}}) d \psi^1 \wedge d \psi^{\bar{1}} \wedge d \psi^2
\wedge d \psi^{\bar{2}} \right),
\ee
where $g_{\tau \bar{\tau}} = -\frac{1}{(\tau - \bar{\tau})^2}$ and
$c_2$, the second Chern class of $\mathcal{M}$, is given by
\be
c_2 = \frac{1}{8 \pi^2} \left( \mathrm{tr}(\mathcal{R} \wedge
\mathcal{R}) - 
\mathrm{tr} \mathcal{R} \wedge \mathrm{tr} \mathcal{R} \right).
\ee
(\ref{indexDensity}) may be rewritten as 
\be
d \mu =  g_{\tau \bar{\tau}} d\tau \wedge d \bar{\tau} \wedge 
d \psi^1 \wedge d \psi^{\bar{1}} \wedge d \psi^2 \wedge d
\psi^{\bar{2}}
 \left[ \epsilon^{ab} \epsilon^{\bar{a}\bar{b}} \left(R^1_{a \bar{a}
  1} R^2_{b \bar{b} 2} - R^1_{a \bar{a} 2} R^2_{b \bar{b} 1}\right) -
\det{g_{a \bar{a}}}\right].
\ee

To evaluate both (\ref{fluxeqns}) and (\ref{indexDensity})
we must obtain a knowledge the periods, which in general is a highly
non-trivial task. However, the manifold defined by
(\ref{manifoldeq}) is of a class that has been extensively
studied. The relevant periods have been computed in
\cite{general,twomod}, following the classic
treatment of the quintic\cite{quintic}, and we will borrow their results.
For the Calabi-Yau described by the hypersurface
\be
P= \sum_{j=0}^4 x_j^{d/k_j} - d \psi x_0 x_1 x_2 x_3 x_4 -
\frac{d}{q_0} \phi x_0^{q_0}x_1^{q_1}x_2^{q_2}x_3^{q_3}x_4^{q_4} = 0,
\ee
the fundamental period in the large $\psi$ region is given by
\be
\label{largepsifund}
\varpi_f(\psi,\phi) = \sum_{l=0}^{\infty} \frac{(q_0 l!) (d \psi)^{-q_0 l} (-1)^l}
{l! \Pi_{i=1}^4 (\frac{k_i}{d}(q_0 - q_i)l)!} u_l(\phi),
\ee
where
\be
u_l(\phi) = (D\phi)^l \sum_{n=0}^{[\frac{l}{D}]} \frac{    l! \Pi_{i=1}^4 (\frac{k_i}{d}(q_0 - q_i)l)!(-D\phi)^{-Dn} }
{ (l-Dn)!n! \Pi_{i=1}^4 (\frac{k_i q_i}{q_0}n + \frac{k_i}{d}(q_0 -
  q_i)l)!} .
\ee
This is obtained by direct integration of $\Omega$ and satisfies the
Picard-Fuchs equation. There are other
independent solutions to the Picard-Fuchs equation having
a logarithimic dependence on $\psi$. In total there are six
independent solutions, one for each 3-cycle, and the actual periods are a linear combination
of these.

The two regions of moduli space that will most interest us are 
the vicinities of the Landau-Ginzburg point $\psi =\phi =0$ and the
the conifold locus $864 \psi^6 + \phi = 1$. To obtain a basis of periods in the small
$\psi$ region, we analytically continue (\ref{largepsifund}) to obtain
\be
\label{smallpsi}
\varpi_f\left(\psi,\phi\right) = -\frac{2}{d} \sum_{n=1}^{\infty}
\frac{\Gamma\left(\frac{2n}{d}\right) 
\left(-d \psi\right)^n u_{-\frac{2n}{d}}\left(\phi\right)}
{\Gamma\left(n\right) \Gamma\left(1-\frac{n}{d}\left(k_1 -1\right)\right) \Gamma\left(1 - \frac{k_2 n}{d}\right)
  \Gamma\left(1-\frac{k_3 n}{d}\right) \Gamma\left(1-\frac{k_4 n}{d}\right)} .
\ee
Here $u_\nu(\phi)$ is related to the hypergeometric functions and is defined through the contour integral
\be
\label{smallpsiu}
u_\nu (\phi) = \frac{2^\nu}{\pi} \int_{-1}^{1} \frac{d \zeta}{\sqrt{1
    - \zeta^2}} (\phi - \zeta)^\nu .
\ee
The contour integral is initially defined for $\textrm{Im}(\phi) > 0$
and then defined over the rest of the plane by deforming the integral
contour. The branch cuts, which are unavoidable when $\nu$ is non-integral,
start at $\pm 1$ and run to $\pm \infty$.

We may derive a basis of periods from the fundamental period in a simple manner. 
If we define
\be
\label{periods}
\varpi_j(\psi,\phi) = \frac{-(2 \pi i)^3}{\psi} \varpi_f(\alpha^j \psi, \alpha^{j q_0} \phi)
\ee
then $\varpi(\psi, \phi) = (\varpi_0(\psi,\phi), \varpi_1(\psi,\phi), 
\varpi_2(\psi,\phi), \varpi_3(\psi,\phi), \varpi_4(\psi,\phi),
\varpi_5(\psi,\phi))$ gives a basis of periods 
known as the Picard-Fuchs basis.
Naively (\ref{periods}) would seem to give 12 independent periods, but there are
interrelations discussed in \cite{general}. The net result is
that, as expected, there are six 
independent periods. This basis is however not symplectic. A symplectic basis is given by
\be
\label{pidef}
\Pi(\psi,\phi)  = m \cdot \varpi(\psi,\phi).
\ee
Here $m$ is computed in \cite{hepth9912147, hepth0312104} and is given by
$$
m = \left( \begin{array}{cccccc} -1 & 1 & 0 & 0 & 0 & 0 \\ \frac{3}{2}
  &  \frac{3}{2} &  \frac{1}{2}
&  \frac{1}{2} & - \frac{1}{2} & - \frac{1}{2} \\
1 & 0 & 1 & 0 & 0 & 0 \\ 1 & 0 & 0 & 0 & 0 & 0 \\ - \frac{1}{2} & 0 &
\frac{1}{2} &  \frac{1}{2} & 0 & 0\\
 \frac{1}{2} &  \frac{1}{2} & - \frac{1}{2} &  \frac{1}{2} & - \frac{1}{2} &  \frac{1}{2} \end{array} \right).
$$

In principle, equation (\ref{pidef}) completely determines the periods
near $\psi = 0$. 
However, it involves the integral expression (\ref{smallpsiu}) for $u_\nu (\phi)$
which is inconvenient for a computational treatment. 
Such a treatment is facilitated by the power series expansion of
$u_\nu(\phi)$ in the small $\phi$ region 
found in \cite{general}. This is given by
\be
\label{usmallphi}
u_\nu\left(\phi\right) = \frac{e^{\frac{i\pi \nu}{2}} 
\Gamma \left(1 + \frac{\nu\left(k_1 - 1\right)}{2}\right)}{2\Gamma\left(-\nu\right)}
\sum_{m=0}^\infty 
\frac{e^{i \pi m/2} \Gamma\left(\frac{m-\nu}{2}\right) \left(2\phi\right)^m}{m! \Gamma\left(1 - \frac{m - \nu k_1}{2}\right)}
\ee

The region of convergence of equations (\ref{smallpsi}),
(\ref{smallpsiu}), and (\ref{usmallphi}) is worth discussing. We will 
here be specific to
the particular Calabi-Yau described in section \ref{Model}. 
This manifold develops a conifold singularity
when $\phi + 864\psi^6 = \pm 1$ and there is also a strong coupling singularity
when $\phi = \pm 1$. The regions of convergence are determined by the singularities.
All three equations are only valid for $|\frac{864 \psi^6}{\phi \pm 1} | < 1$
and equation (\ref{usmallphi}) has the additional restriction $|\phi| < 1$.

We will also be interested in the periods near the conifold locus. As
will be further discussed in section (\ref{coniStatistics}), the periods
here have a certain standard form. However, for their exact numerical
determination, we will use the neolithic approach,
directly evaluating the power series (\ref{pidef})
near the conifold locus.

Finally, the Calabi-Yau we work on is the original manifold $\mathcal{M}$ 
defined by the locus of the polynomial (\ref{manifoldeq}), and not its
mirror $\mathcal{W}$.
$\mathcal{M}$ has a total of 128 complex structure moduli.
We expect some to be removed by the orientifold symmetry, but there
will still be many which we are ignoring. The validity of this was explained in \cite{hepth0312104}.
The group $G$ is a symmetry of (\ref{manifoldeq}) and if we only turn on fluxes invariant under this symmetry
then the superpotential can only have a higher-order dependence on the other moduli. 
It is thus consistent to set all other moduli equal to zero and focus only on
the moduli in equation (\ref{manifoldeq}) and their associated
fluxes. We will comment briefly on the general situation in the last section.

\section{Vacua Statistics}

The two natural regions for testing the Ashok-Douglas formula are the
vicinities of the Landau-Ginzburg point $\psi = \phi = 0$ and the
conifold locus $864 \psi^6 + \phi = 1$.

\subsection{The Vicinity of $\psi = \phi = 0$}
\label{LGpoint}

A symplectic basis for the periods was given in equation
(\ref{pidef}). Let us untangle this in the vicinity of $\psi = 0$. We can
expand $\Pi(\psi, \phi)$ as
\be
\Pi = \underline{a}(\phi) + \underline{b}(\phi) \psi^2 +
\underline{c}(\phi) \psi^4 + \mathcal{O} \left(\psi^6 \right).
\ee
Here $\underline{a}, \underline{b}$ and $\underline{c}$ are vector functions of $\phi$ whose explicit form
arises from the
combination of equations (\ref{smallpsi}), (\ref{usmallphi}), (\ref{periods}) and (\ref{pidef}).
It can be checked that $\underline{a}^\dagger \cdot \Sigma \cdot
\underline{b} = \underline{a}^\dagger
\cdot \Sigma \cdot \underline{c} = 0$, implying
$$
\Pi^\dagger \cdot \Sigma \cdot \Pi = (\underline{a}^\dagger \cdot \Sigma \cdot \underline{a}) +
(\underline{b}^\dagger \cdot \Sigma \cdot \underline{b}) \psi^2
\bar{\psi}^2 + \mathcal{O}\left(|\psi|^6 \right),
$$
and consequently
\bea
K_\phi(\psi, \phi) & = & - \ln \left(-i \Pi^\dagger \cdot \Sigma
\cdot \Pi \right) \nonumber\\
& = & - \ln \left(-i \underline{a}^\dagger
\cdot \Sigma \cdot \underline{a} \right) - \ln \left(1 +
\frac{(\underline{b}^\dagger \cdot \Sigma \cdot \underline{b})}{(\underline{a}^\dagger \cdot \Sigma \cdot
  \underline{a})} \psi^2 \bar{\psi}^2  + \mathcal{O}\left(|\psi|^6\right) \right) \nonumber\\
& = & -\ln\left(-i \underline{a}^\dagger \cdot \Sigma \cdot \underline{a}\right) - 
\frac{(\underline{b}^\dagger \cdot \Sigma \cdot \underline{b})}{(\underline{a}^\dagger \cdot \Sigma \cdot
  \underline{a})} \psi^2 \bar{\psi}^2 + \mathcal{O}\left(|\psi|^6\right).
\eea
Equations (\ref{fluxeqns}) then have the form
\bea
\label{smallphi1}
\frac{1}{\psi} D_\psi W = 0 \Rightarrow &
(f-\tau h) \cdot \left( \underline{\alpha}_1 (\phi) + \underline{\alpha}_2 (\phi) \psi^2 + \underline{\alpha}_3
(\phi) \bar{\psi}^2 \right) & = 0, \\
\label{smallphi2}
D_\phi W = 0 \Rightarrow & 
(f - \tau h) \cdot \left( \underline{\beta}_1 (\phi) +
\underline{\beta}_2 (\phi) \psi^2 \right)
& = 0, \\
\label{smallphi3}
D_\tau W = 0 \Rightarrow & 
(f - \bar{\tau} h) \cdot \left( \underline{a}(\phi) + \underline{b}
(\phi) \psi^2 \right) & = 0,
\eea
where we have dropped terms of $\mathcal{O}\left(|\psi|^4\right)$.
Here $\underline{\alpha}(\phi)$ and $\underline{\beta}(\phi)$ are
complicated functions of $\phi$ depending on the integral
expressions for $u_\nu(\phi)$. However, when $\vert \phi \vert < 1$
the use of the power series expansion in equation
(\ref{usmallphi}) coverts $\underline{\alpha}(\phi)$ and $\underline{\beta}(\phi)$ to a more tractable form.
The leading behaviour of the metric $g_{\alpha \bar{\beta}} =
\partial_\alpha \partial_{\bar{\beta}} K$ is given by
$$
g_{\psi \bar{\psi}} \sim \psi \bar{\psi}, \qquad g_{\psi \bar{\phi}} \sim
\psi \bar{\psi}^2 \bar{\phi}, \qquad g_{\phi \bar{\psi}} \sim
\psi^2 \bar{\psi} \phi, \qquad g_{\phi \bar{\phi}} \sim
1.
$$
This is consistent with our expectations - at the Landau-Ginzburg point $\psi =
\phi = 0$ the metric becomes singular. The curvature 2-form
$\mathcal{R}$ and the Chern class $c_2$ may be
calculated straightforwardly from the full expressions for the metric
using a symbolic algebra program. 
Evaluating the Ashok-Douglas density, we find it has leading behaviour
$$
d \mu \sim g_{\tau \bar{\tau}}\, d \tau \wedge d \bar{\tau} \wedge 
\psi \bar{\psi} d \psi \wedge d \bar{\psi} \wedge
d \phi \wedge d \bar{\phi}.
$$
The regions on which we compare the Ashok-Douglas formula to our
empirical results are balls in $\psi$ and $\phi$ space. We then expect
as leading behaviour
\bea N(\textrm{vacua s.t.}  \quad \vert \psi \vert <
r_1) & \sim & r_1^4, \nonumber\\
N(\textrm{vacua s.t.}  \quad \vert \phi \vert < r_2) & \sim & r_2^2. \nonumber
\eea 

To test this expectation, we generated random fluxes and sought
solutions of equations (\ref{smallphi1}) to (\ref{smallphi3}) using a 
numerical root finder. The range of fluxes used was (-20, 20). This is
not as large as one might prefer. However, a larger range of fluxes
resulted in solutions being produced insufficiently rapidly for our
purposes. In order to be able to trust our truncation of the power
series, we only kept solutions satisfying
\be
\label{solnconds}
\left\vert \frac{864 \psi^6}{\phi \pm 1} \right \vert < 0.5 \textrm{
  and  } \vert \phi \vert < 0.75 .
\ee
When processing the numerical results, there is an important subtlety
  we must account for\footnote{We thank S. Kachru for
  bringing this to our attention.}. It is well known that there is an exact
  $SL(2,\mathbb{Z})$ symmetry of type IIB,
$$
\tau \to \frac{a \tau + b}{c \tau + d}, \qquad 
\left( \begin{array}{c} F_3 \\ H_3 \end{array} \right) \to
\left( \begin{array}{cc} a & b \\ c & d \end{array} \right) \left( \begin{array}{c} F_3 \\ H_3 \end{array} \right).
$$
where $a,b,c,d \in \mathbb{Z}$ and $ad - bc = 1$. Thus each vacuum found has many physically 
equivalent $SL(2, \mathbb{Z})$ copies that we should not double-count.
One way to deal with this would be to fix the gauge explicitly and
then perform the Monte-Carlo analysis. Our approach is instead to
weight each vacuum by the inverse of the number of copies it has
within the sampled flux range. The purpose of this is to ensure that
vacua with many $SL(2, \mbb{Z})$ copies are not
preferred. 

As well as the $SL(2, \mbb{Z})$ symmetry, there is a monodromy near
the Landau-Ginzburg point that needs similar treatment. From the
definition of the periods (\ref{periods}), we can that they
have a monodromy under $(\psi, \phi) \to (\alpha \psi, - \phi)$, where
$\alpha^{12} = 1$, of
$$
\varpi(\psi, \phi) \to a \cdot \varpi(\psi, \phi),
$$
where 
$$
a = \left( \begin{array}{cccccc} 0 & 1 & 0 & 0 & 0 & 0 \\ 0 & 0 & 1 &
  0 & 0 & 0 \\ 0 & 0 & 0 & 1 & 0 & 0 \\ 0 & 0 & 0 & 0 & 1 & 0 \\
0 & 0 & 0 & 0 & 0 & 1 \\ -1 & 0 & 0 & 0 & 0 & 0 \end{array} \right).
$$ In the symplectic basis, the monodromy matrix $A$ is given by $m
\cdot a \cdot m^{-1}$. The effect of this monodromy is to produce a 
family of physically equivalent solutions related by
\bea
(\psi, \phi) & \to & (\alpha^{-1} \psi, - \phi), \nonumber \\
f & \to & f \cdot A, \nonumber \\
h & \to & h \cdot A. 
\eea
When weighting vacua we need to find the total number of copies
lying within the sampled flux range from
all symmetries and monodromies. This has important systematic effects
as vacua with smaller values of $f_i$ and $h_i$, and thus smaller
values of $N_{flux}$, have more copies. A naive counting that
neglects the symmetries or monodromies that are present thus places
undue emphasis on vacua with smaller values of $N_{flux}$.

\begin{figure}[ht]
\linespread{0.2}
\begin{center}
\makebox[10cm]{
\epsfxsize=10cm
\epsfysize=7cm
\epsfbox{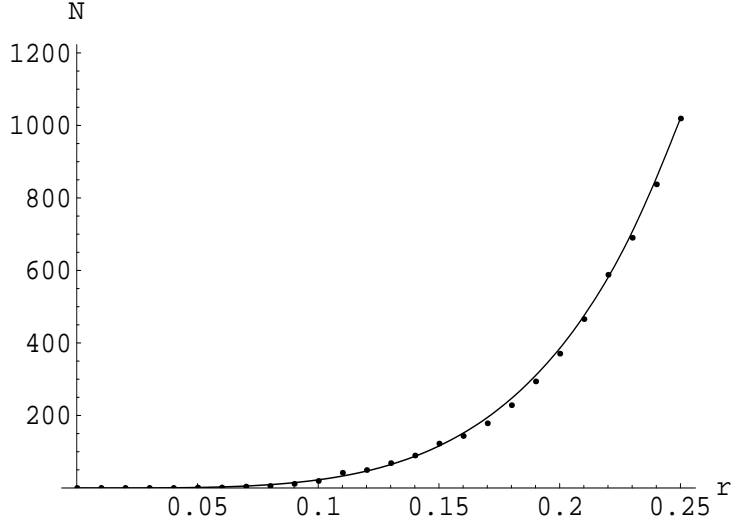}}
\end{center}
\caption{ Number of vacua with $\vert \psi \vert < r$. The value of $N$
  plotted includes a weighting due to the $SL(2,\mathbb{Z})$ copies of
  each vacuum within the range of fluxes sampled. The dots represent
  the numerical results, and the continuous line the
  (rescaled) numerical integration of $\int_{\vert \psi \vert
  < r} d \mu$, where $d \mu$ is the index density. The flux range used
  was (-20,20).}
\label{psismallresults}
\end{figure}

\begin{figure}[ht]
\linespread{0.2}
\begin{center}
\makebox[10cm]{
\epsfxsize=10cm
\epsfysize=7cm
\epsfbox{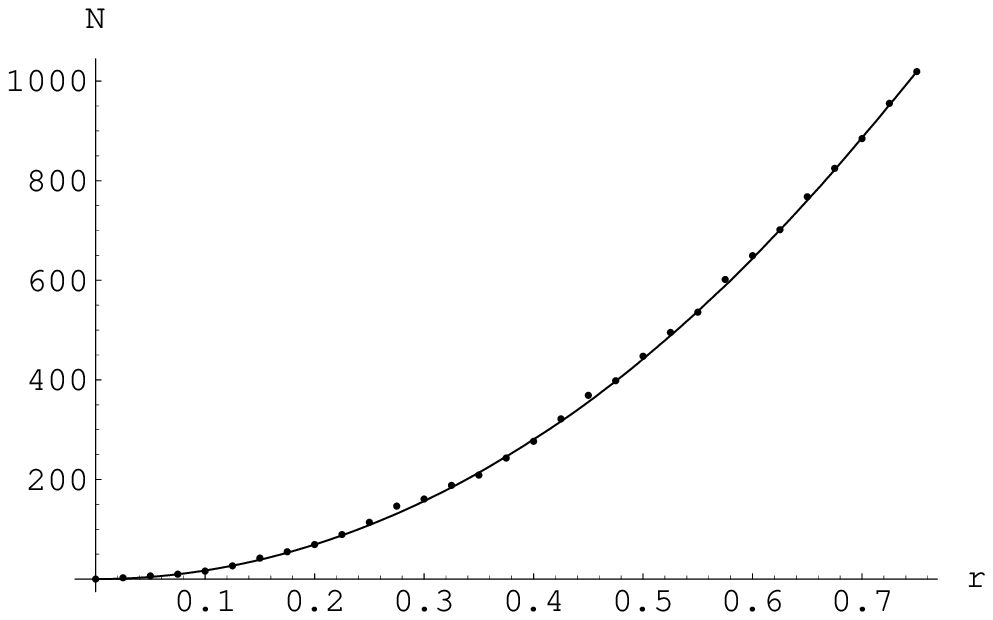}}
\end{center}
\caption{The weighted number of vacua, $N$, with $\vert \phi \vert <
  r$. The dots represent the numerical results and the continuous line the
  numerical integration of $\int_{\vert \phi \vert < r} d
  \mu$, rescaled by the same factor as in diagram \ref{psismallresults}.}
\label{phismallresults}
\end{figure}

\begin{figure}[ht]
\linespread{0.2}
\begin{center}
\makebox[10cm]{
\epsfxsize=10cm
\epsfysize=7cm
\epsfbox{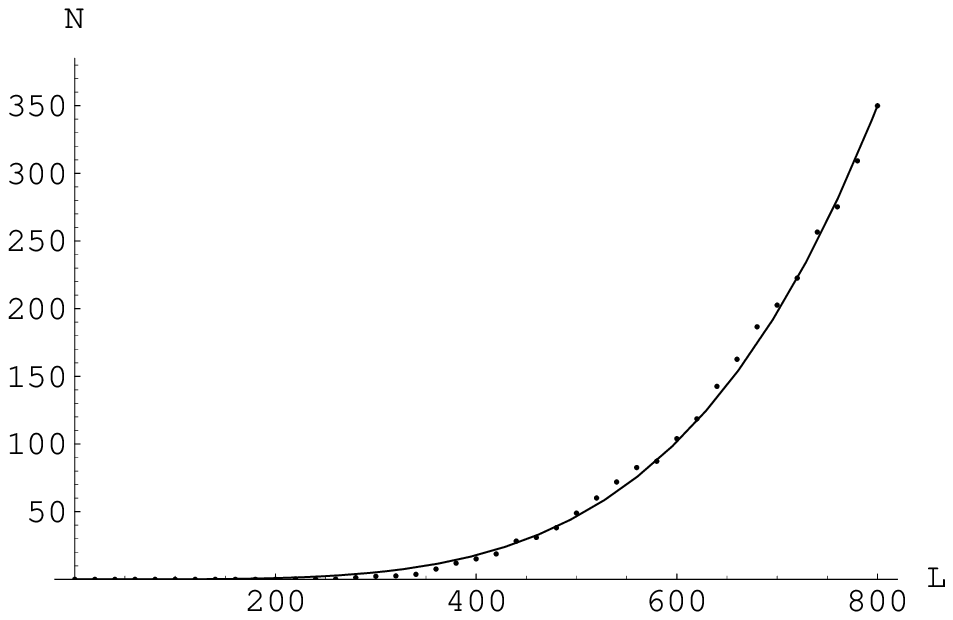}}
\end{center}
\caption{The weighted number of vacua $N$ with
  $N_{flux} < L$. $4.5 \times 10^6$ sets of
  flxues were generated, with values of $L$ equally distributed
  between $0$ and $800$ and the range of fluxes being $(-20,20)$. The
  results are fit by $N \sim L^{4.3}$.}
\label{Ldependence}
\end{figure}

We looked at the distribution of vacua within fixed balls in
$\psi$ and $\phi$ space. 
In figure \ref{psismallresults} we plot the number of vacua satisfying
(\ref{solnconds}) and having $| \psi | < r$. 
The results are seen to agree well with the theoretical prediction.
Likewise, figure \ref{phismallresults} shows the distribution of vacua for a
ball $\vert \phi \vert < r$ in $\phi$ space. The continuous
line again represents the cumulative number of vacua and the dots the
rescaled numerical integration of  $\int_{\vert \phi \vert
  < r} d \mu$. The empirical results are 
again well captured by the theoretical prediction.
Finally, in figure \ref{Ldependence} we examine the dependence of the number of
vacua on the distance in flux space $N_{\mathrm{flux}} = f^T \cdot
\Sigma \cdot h$. The graph is fit by $N \sim L^{4.3}$. This is
surprising, as the expected scaling is $L^6$. Furthermore, in the
vicinity of the Landau-Ginzburg point for an analogous one-modulus
example, the correct $L^4$ scaling is found\cite{skpaper}. As we will
discuss further in section \ref{Discussion}, we believe our results
are an artifact of the small flux range used, and that were a larger
flux range used we would obtain the correct scaling. As we will
shortly describe, in the vicinity of the conifold locus we do obtain
the expected scaling with a flux range of $(-40,40)$.

\begin{figure}[ht]
\linespread{0.2}
\begin{center}
\makebox[8cm]{ \epsfxsize=12cm \epsfysize=7cm
\epsfbox{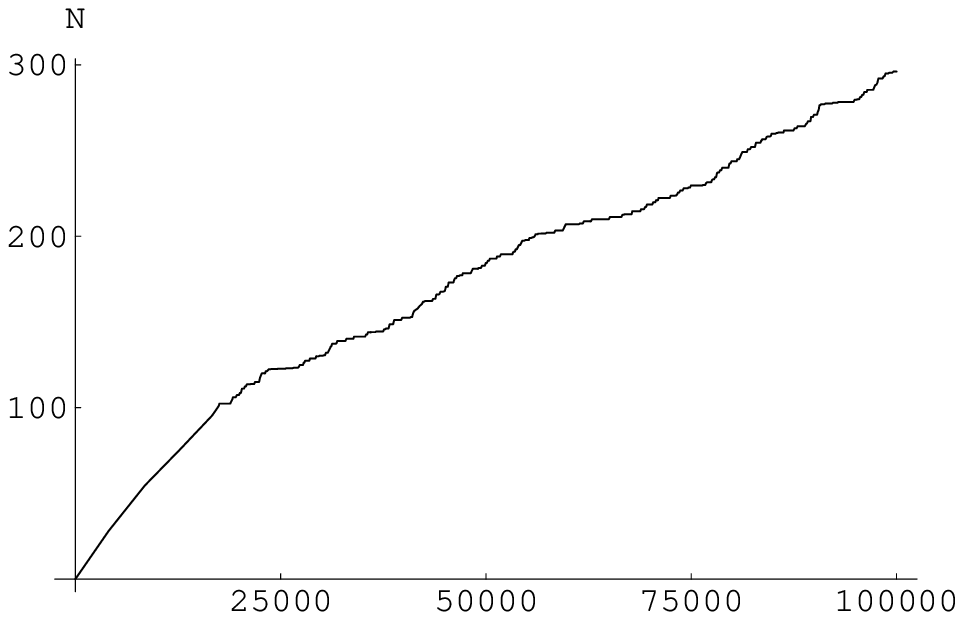}}
\end{center}
\caption{The value of $(2 \pi )^4 e^K |W|^2$ in units of $(\alpha')^2$
  for vacua in the vicinity of $\psi = \phi = 0$, for
  $(2 \pi)^4 e^K |W|^2 < 100000$. The flux range was (-20, 20), and the
  vacua satisfied the conditions (\ref{solnconds}).}
\label{mainSusyScale}
\end{figure}

It is also of interest to study the supersymmetry breaking scale,
as measured by $\alpha'^2 (2 \pi)^4 e^K \vert W \vert^2$, for
vacua in the vicinity of $\psi = \phi = 0$. This is shown in figure
\ref{mainSusyScale}. 
The most noticeable thing about this graph is that the distribution of the susy 
breaking scale is uniform near the origin and that the vast majority of vacua therefore 
have a high supersymmetry breaking scale. We can also observe from the
structure of the
graph that arbitrarily small values of the gravitino mass should exist
as well as supersymmetric solutions. In reference \cite{hepth0312104},
explicit supersymmetric vacua were obtained for this model.

\subsection{The Vicinity of the Conifold Locus}
\label{coniStatistics}

The Calabi-Yau $\mathcal{M}$ has a codimension one conifold degeneration along the moduli space 
locus $864\psi^6 + \phi = 1$. The moduli space curvature diverges
near a conifold point and the expectation is that vacua should cluster
near the conifold locus.

To study these vacua numerically, we must restrict attention to a
small region of the conifold locus where we can compute the periods
explicitly. 
We take this region to be the 
neighbourhood of the point 
$\phi = 0,  \psi = \psi_0 = 864^{-\frac{1}{6}}$. If we write $\psi =
\psi_0 + \xi$ and truncate the periods at first order in $\xi$ and
$\phi$, then $\Pi(\xi, \phi) = (\mathcal{G}_1, \mathcal{G}_2, \mathcal{G}_3, z_1,
z_2, z_3)$ is approximated by
\bea
\mathcal{G}_1 & = & 3202\xi + 171.8 \phi + \mathcal{O}(\xi^2, \phi^2, \xi \phi) \nonumber \\
\mathcal{G}_2 & = & 4323 - i(1553\xi + 107.4 \phi) + \mathcal{O}(\xi^2, \phi^2, \xi \phi)\nonumber \\
\mathcal{G}_3 & = &  (-492.7 + 1976.8i) + (371.0 - 300.2i)\xi + (-259.0
-59.0i)\phi  + \mathcal{O}(\xi^2, \phi^2, \xi \phi) \nonumber \\
z_1 & = &  \frac{-1}{2\pi i}\mathcal{G}_1 \ln \left(\mathcal{G}_1 \right) + 784.8i -
  2306i\xi -44.35i\phi  + \mathcal{O}(\xi^2, \phi^2, \xi \phi) \nonumber \\
z_2 & = & (-994.6 - 184.8i)  + (861.9 + 476.5i)\xi  + (9.91 -
      112.7i)\phi  + \mathcal{O}(\xi^2, \phi^2, \xi \phi) \nonumber \\
z_3 & = &  i(369.5 - 953.0\xi + 225.4\phi) + \mathcal{O}(\xi^2, \phi^2, \xi \phi)
\eea
The numerical values were found by evaluating the series
(\ref{smallpsi}) up to one hundred terms in $\psi$ and twenty-five
terms in $\phi$. The values for the coefficients of the $\mathcal{O}(\xi, \phi)$
terms were sensitive to the number of terms used in the power
series at the level of a couple of percentage points. We did not keep
terms quadratic in $\xi$ and $\phi$ - inclusion of these would lead to 
$\mathcal{O}(\xi)$ corrections to the results below.

The general form of the periods is standard. The cycles
corresponding to $\mathcal{G}_2, \mathcal{G}_3, z_2$ and $z_3$ are all remote from the
conifold singularity, resulting in the associated periods being both regular and
non-vanishing near the conifold degeneration. 
Recalling that the conifold is a cone with a base that is
topologically $S^2 \times S^3$,
the cycle corresponding to $G_1$ is the $S^3$
 that shrinks to zero size along the conifold locus.
 The period along this cycle is regular near the conifold locus and
 vanishing along it. Finally, the cycle corresponding to $z_1$ is that
 dual to the $S^3$. It is therefore not uniquely defined and is in
 fact well known to have a 
monodromy under a loop around the conifold
locus. Its period takes the form
$\left( -\frac{1}{2 \pi i}\mathcal{G}_1 \ln (\mathcal{G}_1) +
\textrm{analytic terms} \right)$. 

It is convenient to define $Z = \left( \frac{3202}{171.8} \right) \xi + \phi$ and rewrite the periods as
\bea
\mathcal{G}_1 & = & 171.8 \, Z, \nonumber \\
\mathcal{G}_2 & = & 4323 + 107.4 i \, Z -3554 i \, \xi, \nonumber \\
\mathcal{G}_3 & = & 784.8 - (259 + 59i)\, Z + (5198 + 799 i) \, \xi, \nonumber \\
z_1 & = &  \frac{-1}{2\pi i}171.8 \, Z \ln \left( Z \right) + 784.8
-44.4i \, Z -1479 i \, \xi, \nonumber \\
z_2 & = & (-994.6 - 184.8i) + (9.91 - 112.7i) \, Z + (677 + 2577 i)\, \xi, \nonumber \\
z_3 & = & i(369.5 + 225.4 \, Z -5154 \, \xi). 
\label{zPeriods}
\eea

Z is a measure of the distance from the conifold locus. We are interested in
vacua extremely close to the conifold locus - typically $\ln{|Z|} < -5$
- and thus we will regard $ |Z| << |\xi| << 1$. Having set up the
periods (\ref{zPeriods}), 
we can now compute the
Ashok-Douglas expectation for the index density and compare this with
numerical results. 

Let us now solve equations (\ref{fluxeqns}). First,
\be
\label{taustabilisation}
D_{\tau} W = 0 \Rightarrow  (f - \bar{\tau}h)\cdot \Pi  =  0
\Rightarrow \tau = \frac{f \cdot \Pi^\dagger}{h \cdot \Pi^\dagger} .\\
\ee
This can be written as 
\be
\label{tauvalue}
\tau = \frac{a_0 + a_1 \bar{\xi}}{b_0 + b_1 \bar{\xi}} + \mathcal{O}(Z
\ln Z),
\ee
where $a_i$ and $b_i$ are flux-dependent quantities.
Next,
\be
\label{xivalue}
D_\xi W = 0 \Rightarrow (f - \tau h)\cdot (c_0 + c_1 \xi + c_2
\bar{\xi} + \mathcal{O}(\xi^2)) = 0.
\ee
Using (\ref{tauvalue}) this becomes a linear equation for $\xi$,
easily solved to determine $\xi$ and $\tau$. We finally need the value
of $\ln Z$. This is obtained by considering
\be
\label{zstabilisation}
D_Z W  = 0 \Rightarrow (f_4 - \tau h_4)\ln Z = (d_0 +d_1 \tau)+
(d_2 + d_3 \tau) \xi + (d_4 + d_5 \tau) \bar{\xi}  
\ee
Substituting in for $\tau$ and $\xi$ from (\ref{tauvalue}) and
(\ref{xivalue}) then gives the value of $\ln Z$.

When analysing the results we must account for the
$SL(2, \mathbb{Z})$ copies discussed in section
\ref{LGpoint}. There is also a monodromy near 
the conifold. When solving for $\ln Z$, we
impose no restriction on the imaginary part of $\ln Z$. There is
then a monodromy 
\bea
\ln Z & \to & \ln Z + 2 \pi i, \nonumber \\
(f_1, f_2, f_3, f_4, f_5, f_6) & \to & (f_1 + f_4, f_2, f_3, f_4, f_5,
f_6), \nonumber \\
(h_1, h_2, h_3, h_4, h_5, h_6) & \to & (h_1 + h_4, h_2, h_3, h_4, h_5,
h_6).
\eea
corresponding to a loop in moduli space around the conifold
locus. This gives a further source of physically equivalent solutions
that should not be double-counted.
For each vacuum, we then count the total number of copies within the
specified flux range and weight by the inverse of this number.

Results are shown below. In figure \ref{coniClustering} 
we show the clustering of vacua by plotting the distribution of vacua transverse
to the conifold locus. 
\begin{figure}[ht]
\linespread{0.2}
\begin{center}
\epsfxsize=0.85\hsize
\epsfbox{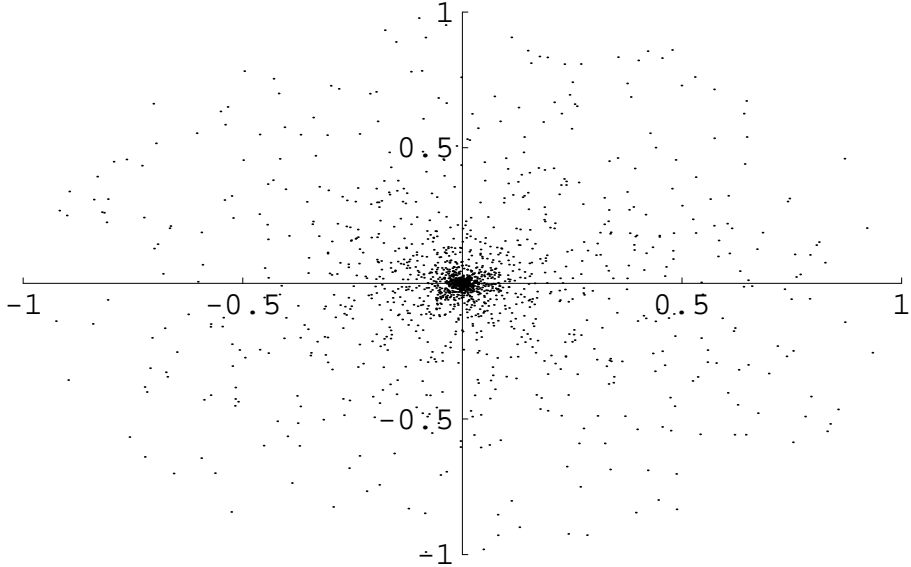}
\end{center}
\caption{The value of $Z$ for vacua near the conifold. We have restricted to
  $|Z| < 0.0001$ and have rescaled the above plot by $10^4$. The flux range used was (-40,40).}
\label{coniClustering}
\end{figure}
A similar plot of vacua parallel to the conifold locus shows no such clustering,
indicating that there is no preferred position along the conifold
locus. In figure \ref{ConiLogVacuaPlot} we perform a quantitative
comparison with the expected vacuum density, finding very good
agreement over a large range of values of $\ln Z$. 
\begin{figure}[ht]
\linespread{0.2}
\begin{center}
\epsfxsize=0.85\hsize
\epsfbox{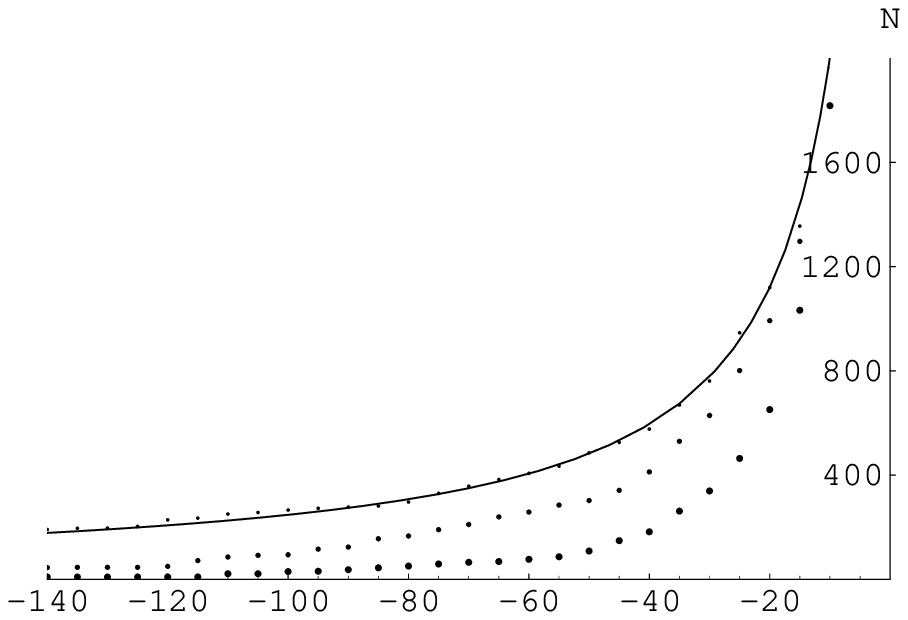}
\end{center}
\caption{The distribution of vacua transverse to the conifold. We plot
  the number of vacua having $\ln Z < D$ for $D \in (-120, -5)$
  against $D$, restricting to $\vert \xi \vert < 0.05$. The dots refer to the
  vacua found numerically and the smooth line to the Ashok-Douglas
  prediction. We include results for three flux ranges - (-20, 20),
  (-30, 30) and (-40, 40). The fit of the results to the expected
  distribution improves markedly as the flux range is increased.} 
\label{ConiLogVacuaPlot}
\end{figure}
In figure 
\ref{ConiFluxVacuaNumbers} we examine the scaling with $L$ of the number of
vacua having $N_{flux} < L$. This reproduces the expected $L^6$
scaling of (\ref{ashokdouglasdensity}).
\begin{figure}[ht]
\linespread{0.2}
\begin{center}
\epsfxsize=0.85\hsize
\epsfbox{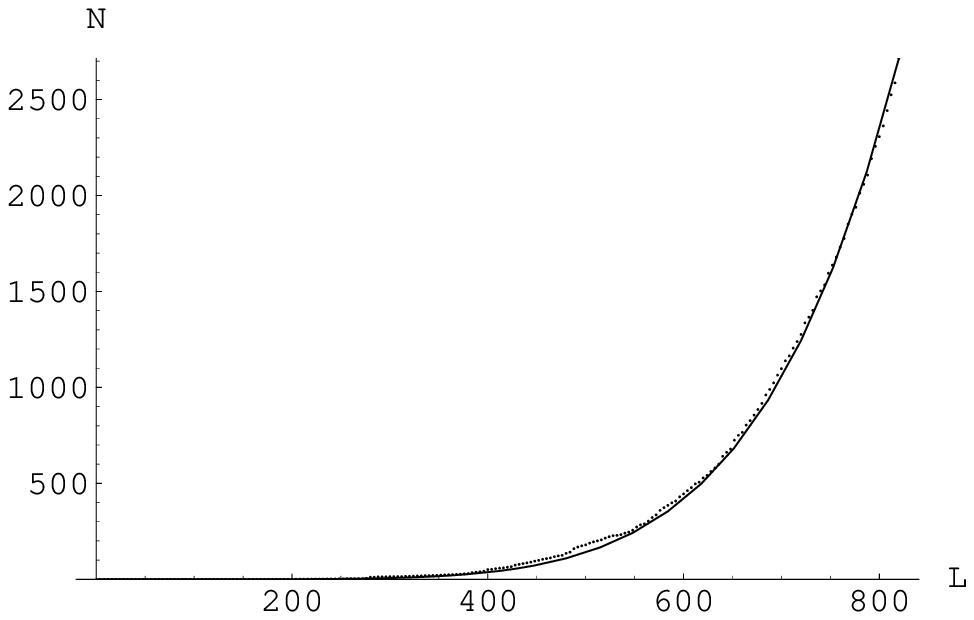}
\end{center}
\caption{The weighted number of vacua with $N_{\mathrm{flux}} < L$. The curve
  is fit well by $N \propto L^{6}$. The range of fluxes used was (-40,40).}
\label{ConiFluxVacuaNumbers}
\end{figure}
We also computed the susy breaking scale, $\alpha '^2 (2 \pi)^4 e^K \vert W \vert ^2$. 
The distribution of susy breaking scales is plotted in figure
\ref{coniSusyScale}, and is uniform near zero.
\begin{figure}[ht]
\linespread{0.2}
\begin{center}
\epsfxsize=0.8\hsize
\epsfbox{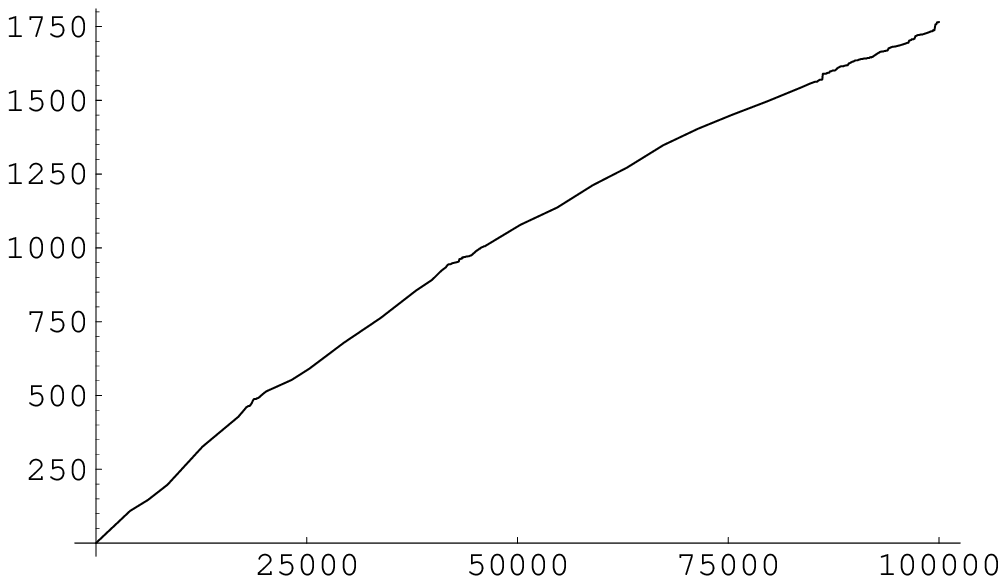}
\end{center}
\caption{The value of $(2 \pi )^4 e^K |W|^2$ in units of $(\alpha')^2$ for vacua near the conifold, for
  $(2 \pi)^4 e^K |W|^2 < 100000$. The flux range was (-40, 40) and we
  restricted to vacua satisfying $\vert \xi \vert < 0.05$.}
\label{coniSusyScale}
\end{figure}

\section{Summary and Discussion}
\label{Discussion}

Let us briefly summarise our results.
\begin{enumerate}
\item We constructed a large class of flux vacua for the two-moduli
  Calabi-Yau threefold $\mbb{P}^4_{[1,1,2,2,6]}$. We independently
  computed the Ashok-Douglas density and compared with our results.
We find good agreement which improves as the range of fluxes is increased.
The number of vacua was limited mostly by working in two patches in
  moduli space: the region near $\psi=\phi=0$ and the region close to
  the conifold singularity $\phi+864 \psi^6=\pm 1$.
\item We found a large concentration of vacua close to the conifold
  singularity as predicted, with the detailed distribution of the vacua being
  in close accordance with expectation.
\item In both regions, the values of the superpotential
  are uniformly distributed near zero. As a consequence, large values of the
  superpotential, corresponding to a large supersymmetry breaking scale, are
  more abundant than small ones. 
\item In both regions the number of models scaled as a power
of the maximum permitted value of $N_{flux}$, $L_{max}$. In the vicinity of the conifold we
reproduced the expected $L^6$ scaling. In the region close to the
Landau-Ginzburg point we only achieved an scaling as $L^{4.3}$. We attribute the failure to
achieve the expected scaling in this region  to the smaller
range of fluxes used there.
\end{enumerate}

An important feature of our analysis is that the flux range used has a
significant effect on the results. If the flux range is insufficiently
large, then the distribution of vacua found numerically will not fit
with the theoretical density. This is most strikingly illustrated for
the case of the conifold in figure \ref{ConiLogVacuaPlot}. Similar
behaviour was seen for the scaling of the number of vacua with $L$ -
in the vicinity of the conifold locus, a reduction of the flux range
to (-20,20) caused a reduction in the power of the scaling from
$\approx 6$ to $\approx 5$. Given this, we believe that the scaling of
$N(\textrm{vacua } \vert N_{flux} < L) \sim L^{4.3}$ found in the 
vicinity of the Landau-Ginzburg point is simply
an artefact of the flux range used.

We can see a similar dependence on the range of fluxes used in the
distribution of the dilaton. After being transformed to the $SL(2,
\mbb{Z})$ fundamental region, the expectation from equation
(\ref{indexDensity})
is that the number of vacua with $\textrm{Im}(\tau) > \tau_0$ should scale
like $\frac{1}{\tau_0}$. In figure 9 we compare this with numerical
results arising from using flux ranges $(-20,20)$, $(-40,40)$ and
$(-60,60)$. 

\begin{figure}[ht]
\label{dilatonPlot}
\linespread{0.2}
\begin{center}
\epsfxsize=0.8\hsize
\epsfbox{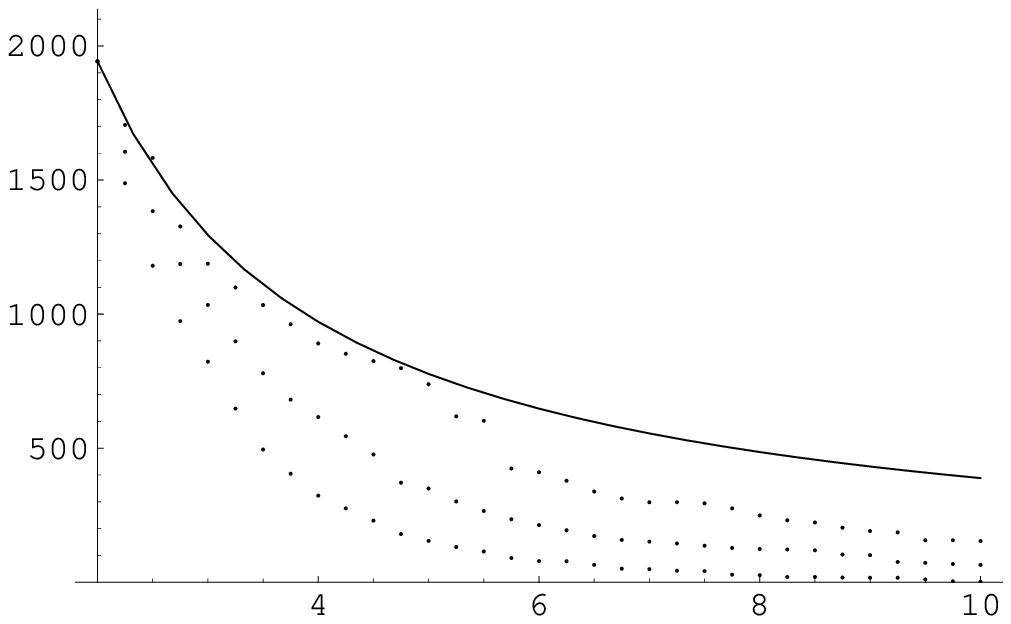}
\end{center}
\caption{Distribution of vacua near the conifold locus with $\textrm{Im}(\tau) > y$ for $y \in
  (2,10)$. Results have been brought to the same scale and 
plotted for three separate flux ranges:  (-20,20), (-40,40) and
  (-60,60). We see that as the flux range increases the empirical plot
  moves closer to the expected result (represented by a smooth line).}
\end{figure}

We see that as the flux range increases the empirical
distribution tends towards the theoretical one, and also that even
with a flux range of $(-60,60)$ the two distributions do not yet fully match. 
In general terms these results are reassuring in the sense that
arbitrarily large values of the dilaton can be obtained, consistent
with the weak coupling aproximation although they are not
statistically preferred.

As we have concentrated on only two complex structure moduli from a
total of 128, we have explored only a minuscule sample of the full set
of flux vacua for this model.
Still, we obtained sufficient statistics to see the expected
distribution of vacua.
There is a clear computational obstruction
to exploring the full spectrum of models for a given Calabi-Yau. 
However, we have seen that the consistency requirements of 
small string coupling and large volume are largely disfavoured
statistically.
The relevant class of vacua will then be suppressed compared to 
the large estimated totals of order $10^{100}$. Nonetheless, given
this large number it is reasonable to expect many solutions with 
very small values of the supersymmetry breaking scale (which
in the KKLT scenario becomes the large volume constraint). It would be
desirable to develop techniques that, rather than simply counting
the number of vacua by numerically scanning the parameter space,
select only those which satisfy the consistency and phenomenological
requirements. Some small values of $\vert W \vert$ can be found
explicitly - \cite{hepth0312104} contains an example where $\vert W
\vert \approx 4 \times 10^{-3}$ and the first reference of
\cite{nonpert} a case where $\vert W \vert ^2 \approx 2 \times 10^{-4}$.
The efficient discovery of phenomenologically acceptable vacua is harder;   
the techniques recently developed in \cite{genetic} may be of
use in this regard. Finally, let us note that in our study of this
model we have addressed neither the
issue of soft supersymmetry breaking terms induced solely by the
fluxes nor $\alpha'$ corrections nor
the non-perturbative superpotential required to
complete the moduli fixing and lift the potential to get de Sitter
space as in the KKLT scenario. For recent progress in theses
directions, \cite{sofsusy,hepth0408054,nonpert} may be consulted.

Let us finish with a general observation.
The Calabi-Yau considered here has $h^{1,1} = 2$ and $h^{2,1} =
128$. We have turned on fluxes along the cycles corresponding to only
two of the 128 complex structure moduli. We would expect that some of
the remaining moduli should be frozen out by the orientifold symmetry,
but it would still be obviously impractical
to attempt either to write down or to solve the moduli stabilisation
equations with all fluxes turned on. However, if we assume that the
Ashok-Douglas density remains valid then we can say something about
the generic situation. Suppose we have K cycles supporting
flux and that - as holds for this and many other F-theory models - 
\be
N_{D3} + N_{flux} = L_{max} \sim 1000.
\ee
The Ashok-Douglas density (\ref{ashokdouglasdensity}) tells us that
\be
N(\textrm{vacua } \vert  N_{flux} < L_*) \sim L_*^K.
\ee
The fraction of vacua having $N_{D3} \ge n$ is then estimated by
\be
\frac{N(\textrm{vacua } \vert  N_{flux} \le L_{max} -n)}{N(\textrm{vacua
  } \vert  N_{flux} \le L_{max})} = 
\frac{(L_{max} - n)^K}{L_{max}^K}.
\ee For $K \sim 200$, then for $n=5$ this
is approximately $\frac{1}{e} \approx 0.36$. 
We then see that despite the large amount of $D3$-brane
charge we have to play with, generic vacua have rather small gauge
groups \footnote{This discussion has to be modified when including the
  D3 brane moduli which, when fixed, will contribute to the statistics
  of vacuum
  degeneracy. Probably arguments using the attractor mechanism towards
  enhanced symmetry points \cite{attractor}
can be used to control this extra
  degeneracy. A more detailed analysis of this issue is beyond the
  scope of this note. We thank E. Dudas and S. Kachru for discussions
  on these points.}. 
  This is appealing from a phenomenological point of view
if the standard model were to live on D3 branes as in the realistic models of
\cite{realistic}.
 Of course, the actual numbers depend on the particular Calabi-Yau,
 the number of 3-cycles surviving the orientifold projection and
the modifications to the Ashok-Douglas density when $L \approx (\textrm{a
  few})K$, but the general conclusion should be broadly unaffected.
Clearly, there is no reason to require that our universe belong to
statistically preferred classes of vacua, but given the discussion
above, it is always useful to know which properties are  generic and which are not.

\vspace{1cm}

\noindent {\bf Acknowledgements} \\ 

We would like to thank M. Douglas, A. Giryavets and particularly
S. Kachru for very useful suggestions, comments and discussions. We
are also grateful for interesting conversations with P. Berglund, X. de la Ossa, A. Font,
D. Grellscheid and K. Suruliz. FQ thanks the Aspen Center for Physics
for hospitality during its summer workshop `Strings and the Real
World', where part of this project was
developed. JC thanks EPSRC for a research
studentship. FQ is partially supported by PPARC and a Royal Society
Wolfson award.

\end{document}